\documentclass[twocolumn,10pt]{article}

\usepackage{authblk}

\usepackage[pdftex]{graphicx}
\usepackage{url}
\usepackage[defblank]{paralist} 

\usepackage{xcolor}

\title{Search on Secondary Attributes in Geo-Distributed Systems}

\author[1,2]{Dimitrios Vasilas}
\author[1]{Marc Shapiro}
\author[2]{Bradley King}
\affil[1]{Inria \& Sorbonne Universit\'es, UPMC Univ Paris 06, LIP6}
\affil[2]{Scality}

\date{}

\begin{document}

\maketitle

In the age of big data, more and more applications need to query and analyse
large volumes of continuously updated data in real-time.
In response, cloud-scale storage systems can  extend their interface that
allows fast lookups on the primary key with the ability to retrieve data based
on non-primary attributes.
However, the need to ingest content rapidly and make it searchable immediately
while supporting low-latency, high-throughput query evaluation, as well as the
geo-distributed nature and weak consistency guarantees of modern storage
systems pose several challenges to the implementation of indexing and search
systems.
We present our early-stage work on the design and implementation of an
indexing and query processing system that enables real-time queries on secondary
attributes of data stored in geo-distributed, weakly consistent storage
systems.

\section{Introduction}

Various object storage systems have arisen over the recent years to meet the
needs of internet-scale applications.
These data stores, including BigTable \cite{Chang:2008:BDS:1365815.1365816},
Amazon's Dynamo \cite{DeCandia:2007:DAH:1323293.1294281}, Cassandra
\cite{Lakshman:2010:CDS:1773912.1773922} and HBase \cite{hbase} among others, 
are able of storing large amounts of rapidly evolving data, while maintaining
high performance.
In order to achieve their scalability, these systems typically expose a simple
GET/PUT API which allows access to data only through their primary key.
Although key-based access to data is scalable, it is only useful when the keys
of objects that need to be located are known.
This shifts the responsibilities of representing secondary attributes,
one-to-many relationships and many-to-many relationships to application
developers, and forces them to either fit their application logic into a
key-value model, or maintain data in one system and relationships in another.
Furthermore, it makes it difficult to implement applications that need to
retrieve data by attributes other than the primary key.
As an example of secondary attributes, consider the case of photograph files
where secondary attributes may include information such as file-size, creation
date, access rights, geolocation information, confidence score for
classification classes, and other user-defined attributes.
The ability to perform queries on secondary attributes can be a preferred way
to access data for these applications, and would be a natural retrieval
mechanism that can complement the usual key-based semantics of object storage
systems.

Indexing secondary keys has been studied extensively in systems offering strong
consistency, especially in the context of traditional database systems.
However, modern geo-distributed storage systems present different challenges to
the implementation of secondary indexes.
These systems replicate data over servers across geo-distributed data centres
(DCs).
Client operations are served by accessing the local replica without
synchronising with other DCs.
This ensures minimal latency and enables the system to remain available under
network partition (AP).
As a result, these systems offer weak consistency guarantees, where reads might
return stale values and writes may conflict.
In this setting, indexing systems need to ingest updates locally at each DC,
propagate index updates in the background, and resolve conflicting index
modifications.

Moreover, an increasing number of applications continuously produce large
amounts of data at high rates.
An example is social media applications, such as Twitter
\cite{Busch:2012:ERS:2310257.2310393}, where millions of users continuously post
new content.
This creates the need for real-time search: secondary attributes of
continuously created content need to be searchable within seconds after
creation.

\subsection{Problem Statement and Challenges}
\label{problem_statement}
The research problem addressed in this thesis is enabling efficient discovery
and retrieval of data stored in large-scale, geo-distributed, weakly consistent
object storage systems.
Our goal is to extend these systems with an efficient and scalable indexing
and query processing system, focused on real-time queries on non-primary
attributes.

In this section, we describe the requirements and challenges involved in
extending a geo-distributed, weakly consistent storage system to support
real-time search on secondary attributes.
These challenges guide our design choices for the implementation of our system.
\\

\textbf{Low-latency, high-throughput query processing.} Users and applications
expect to receive query results with minimal latency.
Moreover, cloud-scale search systems must be able to cope with large query
volumes.
In other words, the query processing system must achieve both low latency and
high throughput.
Maintaining high performance is especially challenging for data stores that
scale to very large amounts of data.
\\

\textbf{Evolving dataset.} In the setting of real-time search, data may be
created and updated at a high rate.
Despite that, users and application expect data to be searchable within a short
amount of time.
The indexing system must therefore ingest and index updates achieving both low
latency and high throughput.
In addition, index update operations should not occur significant overhead to
the latency of source data reads and writes.

Search systems that support real-time queries on evolving datasets, must enable
large volumes of concurrent index reads and writes.
Index structures must track updates incrementally as they occur while at the
same time being accessed to answer queries.
\\

\textbf{Geo-distributed, AP data stores.} Today's large-scale
storage systems replicate data across geo-distributed data centres in order to
avoid network latency and tolerate network partitions.
They implement weak consistency models, where client operations are served by
the local replica, while updates are propagated asynchronously to other
replicas.
As a result, search systems must be able to ingest concurrent updates and serve
queries at multiple DCs without synchronisation across replicas.
Moreover, indexing systems must be able to resolve conflicting modifications to
index structures caused by concurrent updates at different replicas.

\subsection{Data and System Model}
\label{data_system_model}
In an object storage system, objects are composed of an uninterpreted blob of
data (content), accompanied by a set of additional metadata attributes, and are
assigned a globally unique identifier (key).
We model the secondary attributes as a JSON-like object attached to each data
object, consisting of a set key-value pairs of binary (text) or numerical data.
Secondary attributes may consist both of system metadata (content size,
timestamp of last modification, author, access control lists), as well as
custom, user-defined attributes.
This representation resembles the model that Amazon's S3 object storage API
\cite{amazons3} implements.

Applications expect to be able to perform both exact match and range queries,
using multiple secondary attributes, and express queries as logical expressions
using conjunctions and disjunctions.

We model the geo-distributed data store as a set of storage servers grouped in 
geo-distributed data centres.
Data is partitioned among servers within a DC, and fully replicated among DCs.
Read and write requests are served from the DC nearest to the client's location
without contacting remote DCs, while updates are propagated asynchronously
among DCs.

\section{Problem Analysis \& Design Space}

In this section we give an overview of the various design questions that affect
the design of a system that supports real-time queries on secondary attributes
in geo-distributed weakly consistent storage systems.

We perform a detailed analysis of the various aspects of described
problem, and discussing how the problem's requirements affect our design
choices.

\subsection{Index Organisation}

In an object storage system as described in \ref{data_system_model}, clients
can perform queries on secondary attributes by translating them to Get
operations.
However, since there are no index structure to enable fast lookups on secondary
attributes, queries will have to scan the entire dataset to select objects that
match the given query.
Moreover, since data is partitioned among severs using the objects' primary
keys, all servers need to be accessed for each given query, resulting in large
network loads.

The above discussion shows the inherent inability of an indexless system to
efficiently and scalably process queries on secondary attributes.
It is thus evident the need to extend these system with distributed secondary
indexes.
Secondary indexes allow parallel access to different parts of the index and
can improve throughput and scalability.

There are two main approaches to organising a distributed index:

\textbf{Colocation (Local indexes).} One approach is to colocate index
structures on the same servers as the source data.
Systems using this approach \cite{Lakshman:2010:CDS:1773912.1773922} need to
query all servers storing index partitions for each index lookup.
This allows low latency index updates, as it does not require communication
among servers.
However, the lookup cost increases linearly with the number of servers in the
system, limiting the scalability of the system.

\textbf{Independent Partitioning (Global Indexes).} Another approach is to
partition the index independently from the data, so that indexes are not
necessarily located on the same servers as the corresponding data.
Systems that choose this approach \cite{196190, tan2014diff} achieve
constant index lookup latency, and better scalability as their throughput
increases with the addition of servers.
However, supporting range queries with an index that uses independent
partitioning is challenging.
Using the underlying storage to store index entries destroys data locality, as 
storage systems use hashing to shard their data.

\subsection{Inter-DC Index Replication}
\label{inter_dc_replication}

Extending a geo-distributed data store to support queries on secondary
attributes requires distributing secondary indexes in multiple data centres.
Same as source data, indexes need to be fully replicated among DCs, so that
queries can be evaluated locally without need for communication with remote
servers.
Geo-replicated indexes can ingest updates locally, and propagate updates among
DCs in the background.

There are two design choices for propagating index updates among DCs.
One approach is to rely on the underlying data store's write log.
Index replicas at each DC process updates as they are appended to the local
write log, either by operation performed locally or by operation propagated
from other DCs.
When using this approach, indexes need to process each new update that is
appended to the log and issue index update operations when required.
This may not be efficient for attributes that are not frequently updated, since
indexes will waste computations by going through large volumes of updates
but rarely updating their indexed values.

A different approach is to use a mechanism that propagates updates directly
among indexes, without relying on the propagation of source data writes.
This mechanism can be either operation-based, and propagate index update
operations among index replicas, or state-based, in which case it will
replicate the state of index structures.
This approach is more suitable for rarely updated attributes, but incurs
increased network traffic in the case of heavily updated indexes.

Another aspect of a geo-replicated indexing system which ingests updates
locally at multiple DCs and propagates them at the background, is the fact that
concurrent index updates may conflict.
As an example of conflicting index updates, consider the case where an object
\texttt{Obj} is concurrently updated in \texttt{DC1}, and \texttt{DC2}.
In \texttt{DC1}, the attribute \texttt{Attr} is set to the value \texttt{A},
while in \texttt{DC2} \texttt{Attr} is set to the value \texttt{B}.
The index structures of each DC are updated accordingly by adding \texttt{Obj}
to the index entries \texttt{Attr:A} and \texttt{Attr:B} respectively.
After propagating these updates and merging, index structures should converge
to the same state.
Furthermore, the merged index state should reflect the conflict resolution
performed by the storage system, which will choose either value \texttt{A} or
\texttt{B} for \texttt{Attr}, using a strategy such as last-writer-wins.

This issue highlights the need for a conflict resolution mechanism which will
ensure that index replicas converge to the same state even in the presence of
conflicting updates.

\subsection{Index Maintenance}

In traditional database systems, data are expected to be searchable immediately
after being updated, as search queries are a primary mechanism for data
retrieval
Additionally indexes are used internally for other operations such as view
maintenance.
These system thus maintain strong consistency between indexes and base tables
by updating their indexes synchronously, in the critical path of each update.

On the other hand, in the context of web search search, content is not expected
to be available for searching immediately.
Web crawlers periodically crawl web content and build indexes, using batch
operations to achieve high throughput.
There, indexes are eventually consistent with the source data, and depending on
the type of the content indexing delays of minutes, hours of even days might be
acceptable.

In the case of secondary indexes in storage systems, these exists a spectrum of
design choices for index maintenance.
The most write-optimised approach is not to maintain the indexes synchronously,
which can have no write overhead but results in stale search results.
On the other end, the most read-optimised approach is to synchronously update
indexes in place; that is, to keep every index entry up-to-date based on the
latest data updates.
This may be expensive task in a global index scheme, where the creation of a
single new object may involve communication with multiple servers to update
indexes for different secondary attributes.
There are other design choices that fall in the spectrum between these two
approaches, as \cite{tang2015deferred} shows.

\subsubsection{Implications of Asynchronous Index Maintenance}

Maintaining strong consistency between indexes and source data may be
prohibitive for distributed data stores that accept high rates of updates, due
to the overhead in write latency caused by the index maintenance task.

Indexing systems may therefore choose to update their indexes asynchronously.
This can be implemented by a background task which subscribes to the storage
system's log, receives updates when they are appended to the log, and maintains
the indexes.

As result of this approach indexes may lag behind the state of the data store
due to message delays or high load, and not contain the effects of recent
writes.
Moreover, the amount of divergence between index and source data may grow
unboundedly depending on the system's load.

At the same time, applications that perform searches on evolving datasets may
require fresh search results, as data is changing quickly and factors such
as potential profit may depend on the ability to obtain fresh search results.

Therefore, to address the requirements of different applications, a search
system may use an additional mechanism that updates search results with recent
- not yet indexed - writes at query time.
Since this mechanism would require additional computations, it creates a
trade-off between query response time and result freshness.
Applications can trade additional computations for results freshness, and
obtain stale results with low latency, or more fresh results with slower
response time.

Moreover, allowing divergence between the state of indexes and the state of the
data store introduces both false-positives; Indexes may contain old entries of
objects whose attribute values have been updated.
False-positives can be removed at query time by checking query results against
the data store.

\subsection{Multi-Resolution Indexing}

High cardinality attributes that have a large number of distinct values, may
have a negative impact on both indexing and query processing.
This is especially true for system metadata such as last access timestamp and
object size (although object size values can be efficiently represented in an
index by their logarithm) which are stored with high precision by the storage
system.
Storing each distinct value as an index entry greatly increases the size of the
index.
Moreover, queries often require small precision ("objects with size greater
than 1GB", "objects last accessed more that 3 months ago").

An approach that can address this challenge is binning, where each index entry
represents a range of values.
Binning can reduce the index size and improve performance, but it also
introduces false positives, as results of a query may partially belong in a
bin.
In this case, every object contained in the bin need to be checked in order to
remove those that do not satisfy the query condition, a process called
candidate check.
When a candidate check is needed, it usually dominates the query response time.

The impact of candidate checks can be minimised by an efficient placement of
bin boundaries.
An adaptive indexing system may dynamically place and adjust bin boundaries
based on the number of objects contained in each bin and the resolution at
which attributes appear in past queries, with the goal of finding the best
trade-off between index size and candidate checks.

\section{Literature Review}

\subsection{Secondary Indexes in Distributed NoSQL Databases}

Current NoSQL systems have adopted different strategies to support secondary
indexes.
The design choices of these approaches depend on the characteristics of each
data store and the expected workloads.

SLIK \cite{196190} extends RAMCloud \cite{Ousterhout:2015:RSS:2818727.2806887},
a distributed in-memory key-value storage system, to provide secondary indexes.
It achieves scalability by partitioning indexes independently of the source
data.
SLIK deals with potential consistency problems that occur as a result of
indexes and source data being stored in different servers by introducing two
mechanisms;
It (1) uses an ordered write approach which ensures that the lifespan of each
index entry spans that of the corresponding object, and (2) uses objects as the
ground truth to determine the liveness of index entries, by rechecking index
lookup results against the source data.
Additionally, SLIK performs long-running bulk operations such as index
creation, deletion and migration in the background, without blocking normal
operations. 
The system implements secondary indexes as B+ trees and stores them as regular
tables in the underlying key-value store.
 
Diff-Index \cite{tan2014diff} and Hindex \cite{tang2015deferred} both extend 
log-structured key-value stores to support secondary indexes.
Both works focus mainly on the scheduling of the index maintenance operations
in order to improve performance of write operations.
Their design decision are further discussed in Section
\ref{sota_index_data_consistency}.
These systems maintain global indexes and store index entries as regular
key-value pairs in the underlying data stores.
They update their secondary indexes using regular GET/PUT operations offered by
the underlying data stores.

Qader et al. \cite{Qader2015EfficientSA} study the secondary indexing techniques
used in state-of-the-art commercial and research NoSQL databases.
More specifically, they categorise secondary indexes in (1) stand-alone
indexes, where indexing structures are built and maintained and (2) filter
indexes, where there is no separate secondary index structured, but secondary
attribute index information is stored inside the original data blocks.
Stand-alone indexing techniques are further categorised to those that perform
in-place update (i.e. for each write the index structures are accessed, updated
and stored back to disk), and those that perform append-only updates.
The authors implement a number of different secondary indexing techniques on
top of LevelDB \cite{leveldb} and study the trade-offs between different
indexing techniques on various workloads.
This work is mainly focused on a single server instance of LevelDB and does not
consider a distributed setting.

\subsection{Searchable Key-Value Stores}

HyperDex \cite{Escriva:2012:HDS:2342356.2342360} is distributed key-value store that
provides an interface for retrieving objects based on secondary attributes.
It implements this functionality not by using secondary indexes, but through
hyperspace hashing.
Objects are deterministically mapped to coordinates in a multi-dimensional
space in which axes correspond to the objects' secondary attributes.
Each server of the system is responsible for a region of the hyperspace and
stores the objects that fall within this region.
Using this mapping, each search operation is mapped to the hyperspace and the 
servers that need to be contacted are determined.
Additionally, HyperDex addresses consistency issues that potentially arise from
concurrent updates and object relocation due to updates in their secondary
attributes by a replication protocol that orders updates by arranging an
object's replicas into a value-depended chain.

Innesto \cite{6753825}, is another searchable key-value data store, that
supports search on secondary attributes without maintaining indexes.
Innesto supports multi-attribute range search on any number of secondary
attributes.
The system distributes data by spatially partitioning the key space and
maintains a hierarchy of partitions to support efficient range search.
To provide secondary attribute search on a table, Innesto creates search
clones. 
Each clone is a separate copy of the entire table partitioned differently based
on a subset of secondary attributes.
Innesto provides a strong data consistency model by using one-round
transactions to consistently update data and search clones in parallel.

Replex \cite{Tai:2016:RSH:3026959.3026991} is a multi-key datastore that
supports queries against multiple keys.
Replex does not maintain secondary indexes but instead uses a replication
scheme that makes use of a replication unit, which combines the notion of a
replica and an index, called replex.
A replex stores a data table and shards the rows across multiple partitions.
All replexes store the same data, and each one sorts and partitions data by
a different sorting key associated with that replex.
The system uses chain replication to replicate a row to a number of replex
partitions, each of which sorts the row by the replex’s corresponding index.

\subsection{Commercial Distributed NoSQL Databases}

Various commercial NoSQL systems support secondary indexes.
In this section we describe the design choices that some well knonw NoSQL
system make to implement, distribute and maintain their secondary indexes.

Mongodb \cite{mongodb} uses the B-tree data structure to implement secondary
indexes, and updates indexes synchronously for each data update.
Cassandra \cite{Lakshman:2010:CDS:1773912.1773922} co-locates indexes at th
same servers with the source data.
Indexes are implemented as hidden tables in the underlying data store, and are
maintained by a background process.
New index entries are written at the same time as the primary data is updated
and old entries are removed lazily at query time.

DynamoDB \cite{dynamodbindex} enables users to create multiple secondary
indexes on a table, and perform query and scan operations on these indexes.
It supports both global and local secondary indexes, and reflects updates to
indexes synchronously at write time.
A global secondary index allows users to query an entire table across all
partitions, while a local secondary indexes allows users to query over a single
table partition.

RiakKV \cite{riakv} is another distributed NoSQL database that supports
secondary indexing \cite{riakindex}.
Indexes are stored locally for each partition and updated synchronously at
write time.
At query time, the system determines the minimum number of partitions that it
needs to examine to retrieve a full set of results, broadcasts the query to the
selected partitions.

\subsection{Range Queries on Distributed Hash Tables}

Distributed storage systems often rely on Distributed Hash Tables (DHTs) as
building blocks to implement partitioning of their data to multiple servers.
DHTs perform hash partitioning to efficiently map keys to servers.
Although hash partitioning achieves load balancing and scalability, it also
destroys data locality.
It is thus challenging to efficiently extend these systems with global indexes
that support range queries.

The problem of supporting range queries on distributed structured overlays has
been thoroughly studied in the context of structured Peer-to-Peer (P2P)
networks, implemented on top of DHTs.
While DHTs are efficient for keyword search which requires point queries
\cite{Reynolds:2003:EPK:1515915.1515918, gnawali2002keyword}, range queries are more
challenging to implement.

\textbf{Over-DHT approaches.} A class of solutions aims at building indexing
structures using the DHT as a building block.
These approaches focus on implementing distributed prefix trees
\cite{Ramabhadran:PHT} and distributed binary trees \cite{1348114}.
Distributed prefix trees create data locality over DHTs by partitioning the
data domain with the use of prefixes.
This strategy provides a global knowledge of the tree structure.
Binary trees, on the other hand, provide a flexible ways to partition the value
space.
The space can be partitioned in equal parts to provide random access to tree
nodes, or using other schemes in favour of load balancing.

The Prefix Hash Tree (PHT) \cite{Ramabhadran:PHT} is a distributed data
structure that enables one-dimensional range queries over any DHT.
PHT creates data locality by using a using a prefix rule to recursively divide
the space of binary keys, forming a binary trie.
Tree nodes are mapped to DHT nodes by computing a hash over the PHT node label
The mapping between PHT nodes and DHT nodes is generated by computing a hash
over the PHT node label.
Looking up a key consists of finding a leaf node whose label is a prefix of
that key.
A range query consists of contacting all leaf nodes whose label fall within the
given range.

Range Search Tree (RST) \cite{1348114} presents an adaptive
protocol to support range queries in DHT-based systems.
The RST data structure is a complete and balanced binary tree where each node
represents a different range.
Each non-leaf node corresponds to the union of its two children, while leaf
nodes correspond to the smallest sub-ranges.
In RST, a set of DHT nodes share the load of each sub-range to improve load
balancing.
RST uses a dynamic mechanism to apply insertions only to a set of sub-ranges
that is needed, based on the query ranges and the load information, instead of
applying insertions to every level of the RST.
Moreover, RST adaptively uses nodes only if their presence in the RST can lower
the overall query cost and optimises itself based on load changes.

\textbf{DHT-dependent approaches.} Another class of solutions aims to adapt
the DHT to support range queries instead of using it as a building block.

MAAN \cite{Cai:2003:MMA:951948.952051} extends Chord with locality preserving
hashing to create data locality and support multi-attribute range queries.
A range query for the interval starts at the node responsible for the lower
bound and traverses successor links until the upper bound is reached. 
A drawback of this approach is that locality preserving hashing provides poor
load balancing under skewed distributions.

Saturn \cite{5677527} uses order preserving functions to support
range queries, and focuses on addressing the challenge of load balancing by
introducing a mechanism for replica placement under skewed distributions.
The system detects overloaded nodes and randomly distributes their load using a
multiple ring architecture, where overloaded nodes replicate their data on a
new ring, using a multi-ring hash function.
Saturn is implemented on top of an order-preserving DHT system such as MAAN.

\subsection{Consistency between Index and Source Data}
\label{sota_index_data_consistency}

Diff-Index \cite{tan2014diff} presents an approach to add secondary indexes in
HBase \cite{hbase}, a distributed LSM store.
The authors show that the characteristics of LSM stores (no in-place update,
asymmetric read/write latency) as well as the distributed nature of the system
make the task of maintaining a fully consistent index with reasonable update
performance particularly challenging.
Diff-index offers different levels of consistency between indexes and source
data, and makes trade-offs between index update latency and consistency.
In particular, the system offers multiple levels of consistency varying from
causal to eventual consistency.
The consistency level can be chosen per index depending on the workload and
consistency requirements.
To implement eventual consistency, Diff-index maintains an in-memory queue
which caches all writes that require index processing.
Writes are immediately acknowledged to the clients, while the index is
maintained by a background process.
Additionally, the system implements session consistency by tracking additional
state in the client library.

Hindex \cite{tang2015deferred} addresses the problem of supporting secondary
indexes on top of log-structured key-value stores with the goal of providing
value-based access to data.
Hindex uses performance-aware approach which decomposes the task of index
maintenance to two sub-tasks, (1) index-insert; inserting new index entries and
(2) index-repair; removing old index entries, and executes the inexpensive
index-insert task synchronously while deferring the expensive index-repair.
The authors propose two scheduling schemes for the index-repair operations;
an offline repair that is coupled with the key-value store's compaction
mechanism, and an online repair where index-repair operations are piggybacked
in the execution path of value-based read operation.

Earlybird \cite{Busch:2012:ERS:2310257.2310393} is the retrieval engine that
lies at the core of Twitter's real-time search service.
Twitter users collectively post over 250 million tweets per day, and Earlybird
achieves to make tweets searchable within 10 seconds after creation.
In order to support the demands of real-time search, the system organises
inverted indexes in two levels: an read-only optimised index and an active
write-friendly, block-allocated index that supports both rapid tweet indexing
and query evaluation.
Moreover, authors present a single-writer, multiple-reader lock-free algorithm
that enforces consistency using a simple memory barrier.

\subsection{Multi-Resolution Indexing}

Binning approaches have been proposed in the context of bitmap indexing, as a
way to reduce storage overhead and improve performance of bitmap indexes on
high-cardinality attributes.
There are two conventional binning strategies for bitmap indexing, equi-width
and equi-depth.
Equi-width binning divides the entire dataset value domain into equal
intervals, while equi-depth binning ensures that each bin contains
approximately an equal number of indexed elements.

A dynamic bin expansion and contraction approach for highly skewed data has
been presented in \cite{Wu:1998:RBI:645980.674285}.
The proposed approached initially constructs bins using the equi-width method.
When a bin grows beyond a threshold, it is expanded into multiple smaller-range
bins.
The authors show that the performance of dynamic expansion approach is
comparable with the optimal partition approach, especially for highly skewed
data.

The work in \cite{Sinha:2007:MBI:1272743.1272746} presents a multi-resolution
bitmap indexing framework designed for use with scientific data.
The authors provide a formal framework for analysing the relationship between
storage and performance of multi-resolution bitmap indexes, deciding the number
of resolutions and size of bins at each resolution, and provide an algorithm
for querying a multi-resolution bitmap index with an arbitrary number of
levels.

\subsection{MapReduce Indexing}

In the context of information retrieval, McCreadie et al.
\cite{MCCREADIE2012873} contribute a step towards understanding the
benefits of indexing large Web corpora using the MapReduce processing paradigm.
The authors describe and evaluate existing methods of performing document
indexing in MapReduce, and propose a novel indexing strategy, optimised for
large Web corpora.
They conclude that early MapReduce indexing techniques, proposed in the
original, MapReduce paper generate too much intermediate map data causing
overall slowness, and therefore these strategies are impractical for indexing
at large scale.
On the other side the proposed indexing strategy scales well with both corpus
size and horizontal hardware scaling.

While the MapReduce paradigm can be efficiently used for performing batch
indexing jobs on large-scale datasets, it is less suitable for incrementally
updating indexes in the presence of high rates of updates.
However, these techniques can be useful for performing batch index creation on
pre-existing datasets, or querying attributes with no existing indexes.

\section{Proposed Approach}

\subsection{Overview}

In this section we present our early stage work on the design of a system that
extends geo-distributed object stores to support search on secondary
attributes.
We describe the mechanisms used for various aspects of our system, and discuss
how these mechanisms enable our system to efficiently address the challenges
discussed so far.
Our next steps include defining the algorithms and policies that will make use
of these mechanisms to efficiently implement the system's functionality,
implementing a prototype of the system, as well as performing experiments to
validate the efficiency of our proposed solution.

Our system enables multi-dimensional queries that use both exact match and
range predicates, as well as logical operators.
It is able to ingest writes and queries concurrently in multiple data centres,
and fully replicates secondary indexes among DCs.
Additionally, it allows clients to specify a bound on search result staleness
of each query, enabling clients to make a trade-off between query response time
and result freshness.

Moreover, our design enables the system to adaptively adjust to the system's
workload.
We describe how our system can be extended to dynamically adjust index
resolution based on attribute value skewness and query distribution, and how it
can adaptively allocate computation resources available for indexing and query
processing in order to cope with high loads.

\subsection{Index Organisation}
\label{index_organization}

We model our system as a network of interconnected logical computation units,
called Query Processing Units (QPU).
Each QPU is responsible for serving a particular set of queries.
Queries posed to the system are processed by being routed through the network
of QPUs.
Moreover, write operations performed in the storage system are also propagated
and indexed using the QPU network.

The network is organised using three types of connections between QPUs, which
express different aspects of the described problem:
\begin{itemize}
  \item  Each QPU is responsible for serving queries for a range of values of
  some secondary attributes.
  A QPU can be connected with other QPUs that are responsible for sub-ranges
  of its range of attribute values (Section \ref{value_space_partitioning}).
  \item Each QPU is responsible for responding to queries with results that
  contain the effects of write operations performed in specified time interval.
  A QPU can be connected with other QPUs responsible for sub-intervals of its
  time interval
  (Section \ref{freshness_interval_partitioning}).
  \item Each QPU is responsible for returning results for data stored in a
  specified set of data centres.
  A QPU can be connected with other QPUs responsible for a subset of DCs
  (Section \ref{data_centre_partitioning}).
\end{itemize}

Furthermore, each QPU maintains a multi-level cache of query results that can
be used to respond to queries (Section \ref{result_caching}).

\subsubsection{Value Space Partitioning}
\label{value_space_partitioning}

\begin{figure}[!b]
  \includegraphics[width=\columnwidth]{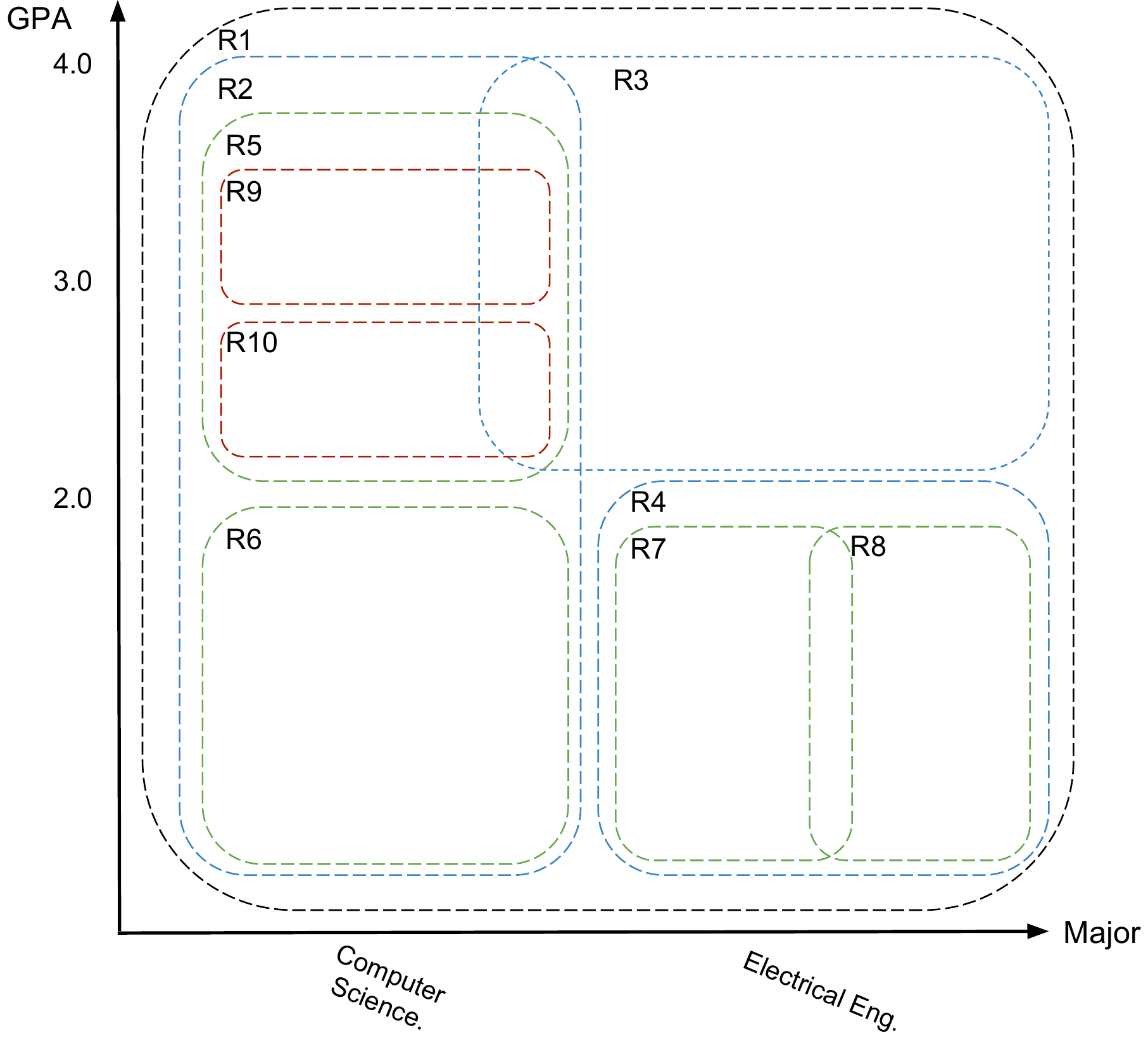}
  \caption{An example of value space partitioning in a system indexing
  student's \texttt{GPAs} and \texttt{Majors}.
  The indexed attributes form a two-dimensional space that is partitioned
  among QPUs.
  A QPU is is responsible for the entire space \texttt{R1}, which is
  sub-divided into regions \texttt{R2}, \texttt{R3} and \texttt{R4}, where
  \texttt{R1} and \texttt{R2} are overlapping.
  \texttt{R2} then then further sub-divided into regions \texttt{R5} and
  \texttt{R6} and so on.
  }
  \label{qpu_2d_space}
\end{figure}

\begin{figure*}[t]
  \begin{centering}
  \includegraphics[scale=.5]{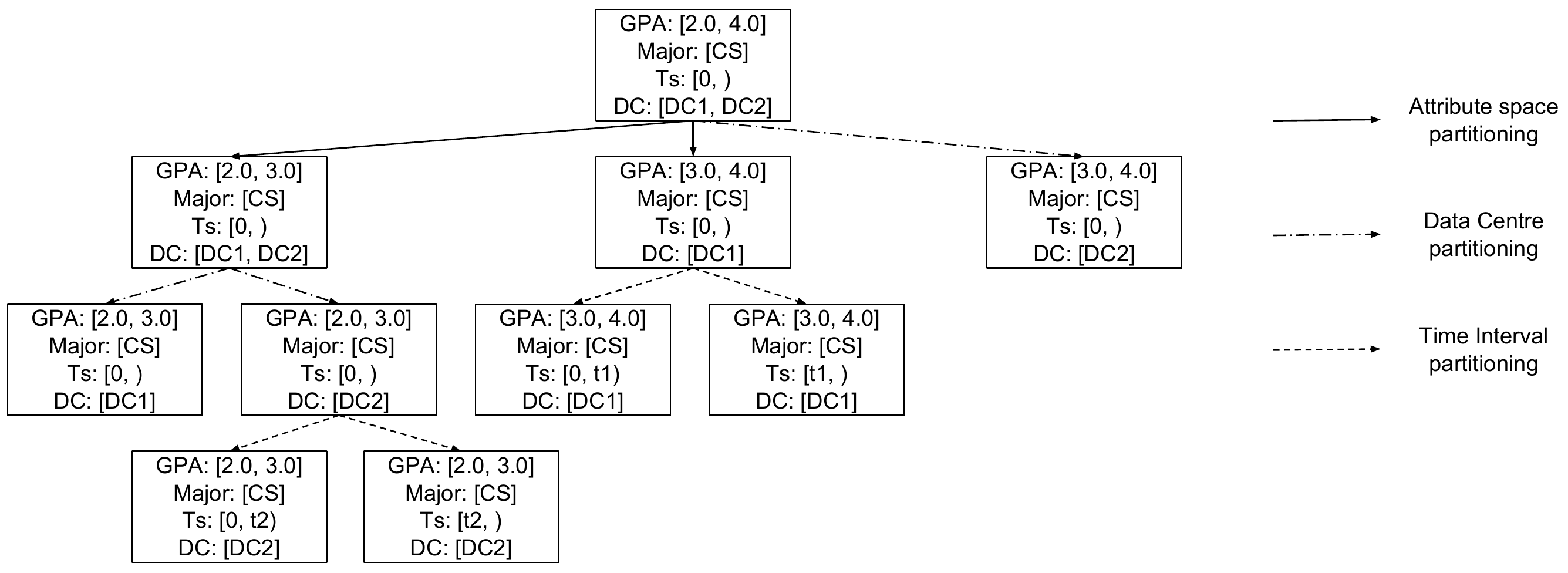}
  \caption{An example of a distributed QPU network for the region
  \texttt{R5} of Figure \ref{qpu_2d_space}.
  The root QPU partitions the attribute value space to two sub-regions.
  Then, QPUs first partition the DC space, by each being responsible for a
  single DC.
  Finally, QPUs partition the freshness interval space, by each being
  responsible for either older or more recent write operations.
  }
  \label{qpu_network_eg}
  \end{centering}
\end{figure*}

Our system uses a static schema for supporting search on secondary attributes.
This schema can be described as set of secondary attributes
Attr\textsubscript{1}, Attr\textsubscript{2}, ..., Attr\textsubscript{N}.
The values of each attribute Attr\textsubscript{i} are ordered and belong in
the range [Min\textsubscript{i}, Max\textsubscript{i}].
Secondary attribute values form a N-dimensional space, where each axis
corresponds to an attribute.
Objects stored in the data store are logically represented as N-dimensional
points in this space, specified by their secondary attribute values.

Each QPU is responsible for a range of values [L\textsubscript{i},
U\textsubscript{i}] of each attribute Attr\textsubscript{i} and acts as a 
service that processes queries in this region.
Both QPUs and queries can be represented as logical N-dimensional rectangles.
QPUs serve queries that intersect with their region of the hyperspace.

Using value space partitioning, QPUs are organised hierarchically as a
distributed R-tree
\cite{Zhang:2009:EMI:1651263.1651267, Sellis:1987:RDI:645914.671636}.
A QPU can be connected to other QPUs that cover smaller sub-spaces
of its region of the hyperspace.
Value space partitioning is not strict, as QPUs with the same parent can
overlap in parts of their regions.
Using the hierarchical structure of the R-tree, QPUs can offload parts of their
computations to other QPUs and then combine retrieved results.

Figure \ref{qpu_2d_space} shows an example of a two dimensional space formed by
a two indexed attributes, which is partitioned into hierarchy a hierarchy of
regions.

\subsubsection{Freshness Interval Partitioning}
\label{freshness_interval_partitioning}

An additional, internal, dimension in the indexing schema is formed by write
operation timestamps.
Each QPU is responsible for returning results that contain effects of write
operations performed in the time interval 
[T\textsuperscript{start}\textsubscript{i},
T\textsuperscript{end}\textsubscript{i}].
Representing result freshness as an additional axis the multi-dimensional
space is a generalization of the distributed R-tree structure.

For a given region of the attribute value space, a QPU can be responsible for
older, already indexed updates, and another QPU can responsible for recent
updates that have not yet been processed.
QPUs responsible for older updates respond to queries by performing index
lookups, while QPUs responsible for newer updates need to pull and process
updates at query time, or scan the underlying data store.
Depending on the freshness requirement of each query, the query is processed by
a combination of QPUs with different freshness intervals.

\subsubsection{Data Centre Partitioning}
\label{data_centre_partitioning}

\begin{figure*}[t]
  \begin{centering}
  \includegraphics[scale=.5]{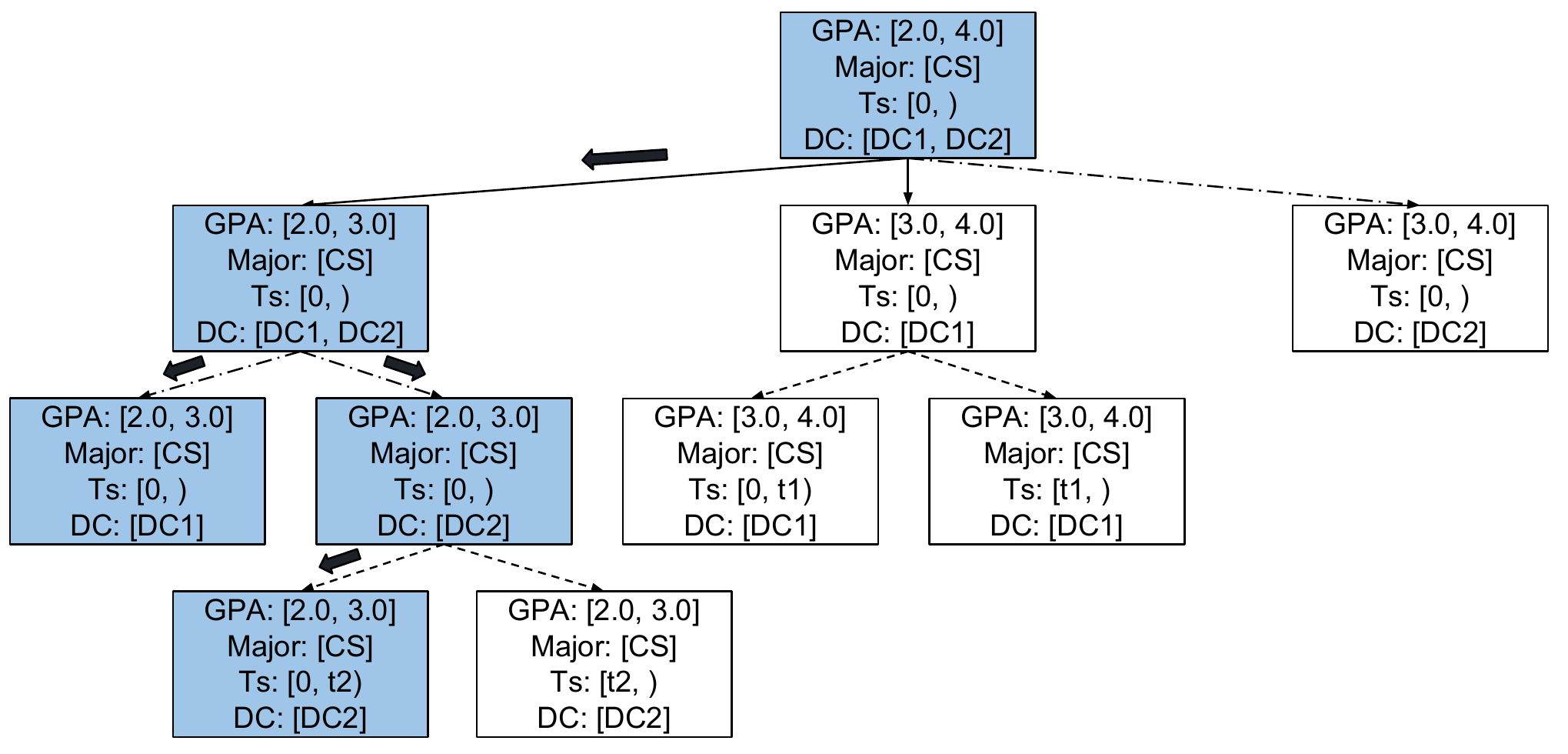}
  \caption{The process of routing the query \texttt{"objects where (GPA>2.0 AND
  GPA<3.0) AND Major=Computer Science"} and \texttt{Result Freshness<t2} posed
  in \texttt{DC1} through the QPU network of the example in Figure
  \ref{qpu_network_eg}.
  }
  \label{qpu_query_processing_eg}
  \end{centering}
\end{figure*}

We further generalise the QPU network by making each QPU responsible for
serving queries using data stored in a set of data centres.
This mechanism enables the QPU network to index write operations and respond to
queries in a geo-distributed multi-DC system.
A QPU responsible for multiple DCs can be connected to other QPUs responsible
for a single DC, and forward query computation to the corresponding QPUs based
on the DC where a given query originates from.

This mechanism enable the geo-distribution of the QPU network, as different
QPUs can be located in different DCs, and is complementary to inter-DC index
replication.
Index updates are propagated among DCs using the mechanism described in Section
\ref{inter_dc_index_replication}.
Using DC partitioning, queries can be processed in a geo-distributed system
even without the use of inter-DC index replication mechanism or in the case of
partial replication, where datasets are not fully replicated among DCs.

Figure \ref{qpu_network_eg} depicts an example of a QPU network containing all
types of connections between QPUs.

\subsubsection{Result Caching}
\label{result_caching}
QPUs can maintain multiple levels of caches storing query results, in order to
enable low latency query processing.
Caches can be considered as partial indexes, containing only a selected subset
of index entries.
Each cache maintains a subset of the index entries stored in lower level
caches.

Internal QPUs of the network can respond to queries using their caches, or
forward the processing to other QPUs they are connected to.
QPUs with no connections to other QPUs maintain full indexes of their indexed
attributes, which are updated based on write operations.
Cache maintenance can be either push-based, where index entries are propagated
from full indexes to caches then pushed to higher cache levels, or pull-based,
where higher level caches request index entries from the lower levels and
eventually from full indexes.

\subsection{Query Processing}

Our system processes queries by routing them through the distributed QPU
network.
Using the network structure, a given query is decomposed into more fine-grained
sub-queries and their computation is assigned to the corresponding QPUs.
Partial results returned from these QPUs are then incrementally combined to
calculate the final response the given query.

Query routing is a recursive process.
Each QPU processes given sub-queries independently using a greedy algorithms.
Given a query, a QPU first determines if it can retrieve the response from its
cache hierarchy.
If the query response cannot be retrieved by a cache, then the next level cache
is visited.
the QPU examines the
other QPUs it is connected to, calculates an efficient decomposition of the
query to sub-queries, and forwards these sub-queries to the corresponding QPUs.

Depending on the type of connections between QPUs, different strategies are used
to determining how queries can be processed:

\textbf{Value space partitioning. } For connections that perform value space
partitioning, the QPU calculates the mapping of a given query to the
N-dimensional space, determines which QPU sub-regions intersect with the
mapping of the given query, decomposes it into sub-queries, and forwards it to
the corresponding QPUs.

\textbf{Freshness Interval Partitioning. } For connections that perform
freshness interval partitioning, the QPU determines which QPUs intersect with
the staleness requirements of the given query, and forwards it accordingly.

\textbf{Data Centre Partitioning. } For DC partitioning connections, the QPU
first determines if index updates are replicated to the DC where the query
originated from.
If index updates are replicated, then the QPU only forwards the query to QPUs
responsible for this DC.
Otherwise, the query is forwarded to other QPUs according to which DCs results
need to be fetched from.

This process is recursively repeated at each QPU, until the given query is
answered by QPU caches or QPUs with no further connections are reached
(leaf QPUs).
Leaf QPUs process sub-queries in parallel, and return lists of objects that
satisfy the given sub-queries to the higher levels.
QPUs then recursively combine the retrieved partial results to calculate the
final query response.

Figure \ref{qpu_query_processing_eg} illustrates the process of decomposing and
routing a query through the QPU network of Figure \ref{qpu_network_eg}.
 
\subsection{Index Maintenance}

Within each data centre, leaf query processing units maintain their indexes
structures asynchronously, in a per-operation basis.
When a write operation is performed locally, it is processed, a new entry is
appended to the storage system's log, and then it is acknowledged to the
client.
The log of the data store exposes a publish-subscribe mechanism that allows
QPUs to receive and process write operations.
Each QPU independently receives all write operations performed in the local DC,
filters operations involving secondary attributes for which it is responsible,
and inserts new index entries or remove deprecated ones accordingly.

\subsubsection{Cache Maintenance}

QPU caches are maintained using a combination of pull and push strategies.
Full indexes keep track of the index entries stored in the lowest level caches.
Once an index entry is updated as a result of a write operation, the index
pushes this entry to the corresponding caches in order to update their outdated
index entries.
Each cache then further pushes updated index entries to higher level caches
when necessary.

Conversely, caches store results of expected queries by pulling them from lower
level caches, and eventually from full indexes.

\subsubsection{Inter-DC Index Replication}
\label{inter_dc_index_replication}

As discussed in Section \ref{data_centre_partitioning}, the QPU network is
geo-distributed among data centres.
QPUs are replicated and different QPUs are responsible for the same regions of
the multi-dimensional attribute space in different DCs.
Queries can therefore be answered by combining QPUs from each DC.
However, this approach incurs additional overhead as it requires communication
between DCs.
We use an additional mechanism that replicates QPU indexes among DCs and
ensures that replicated indexes eventually converge, so that there is no need
for inter-DC communication for query processing.

\begin{figure}
  \includegraphics[width=\columnwidth]{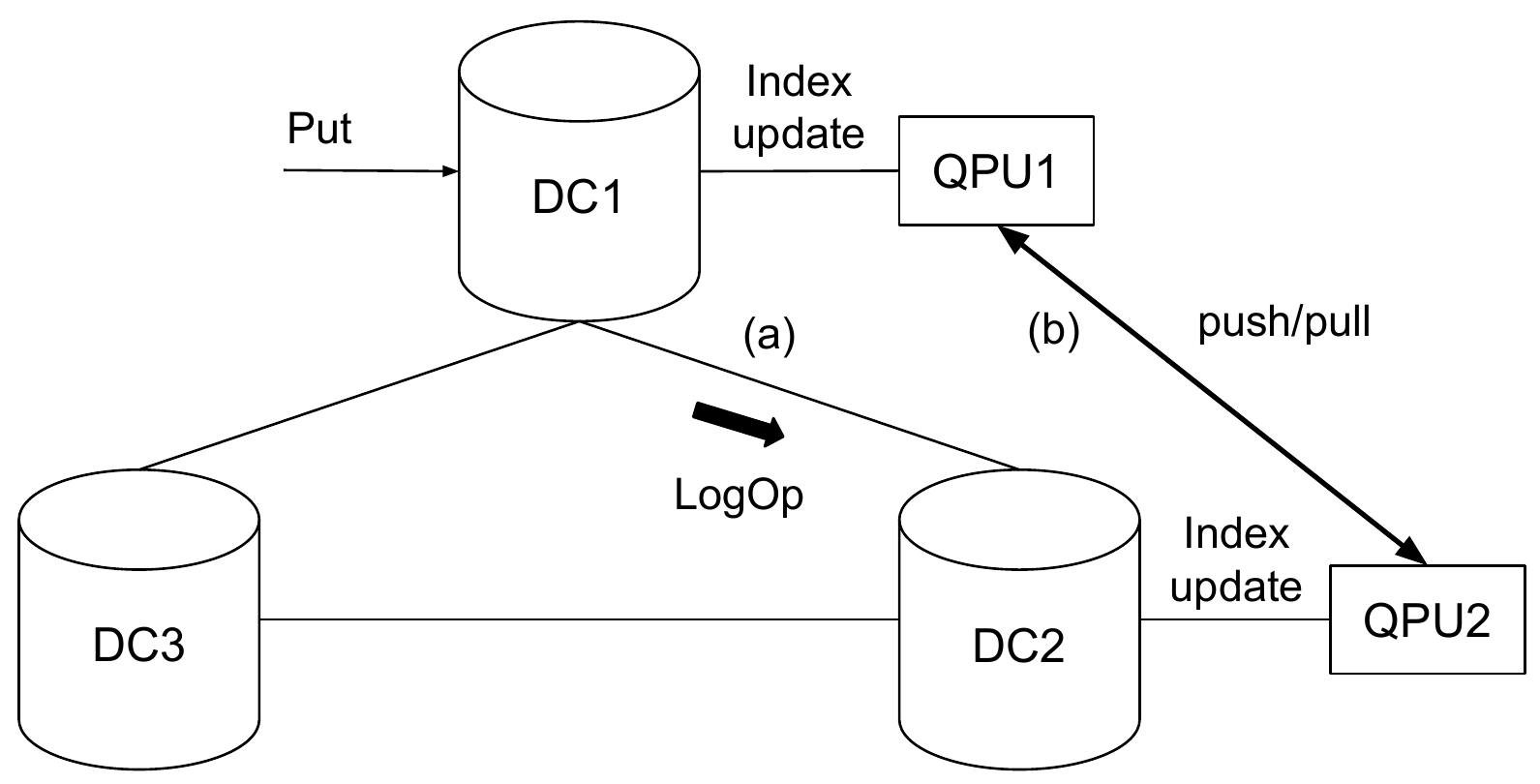}
  \caption{Inter-DC index replication.
  Replicated QPUs are maintained either by receiving write operations that are
  propagated through the data store's log (a) or by directly propagating index
  update operations among QPUs (b).}
  \label{inter_DC_replication}
\end{figure}

Each QPU is maintained locally, and updates are asynchronously propagated among
DCs.
We use two different strategies to manage index replication among DCs, which
are illustrated in Figure \ref{inter_DC_replication}.
One strategy is to replicate log entries of source data writes.
QPUs receive all local write operations as well as writes propagated from other
DCs, and use the information provided by log entries to maintain their indexing
structures.
The second strategy involves direct communication among QPUs.
QPUs receive and process only local writes, and asynchronously propagate index
update operations to other DCs.
Alternatively, QPUs can be updated by pulling the state of corresponding QPUs
from other DCs and merging it with their own.

Each of these strategies is suitable for indexes with different
characteristics.
When indexing attributes that are rarely updated, processing all write
operations, local and remote, is inefficient and may result in waste of
computation resources.
In these cases, propagating index operations is more efficient.
On the other hand, in cases where an indexed attribute is frequently updated,
propagating a large volume of index operations may add unnecessary additional
network traffic to the system.

Since each query processing unit is maintained independently, the indexing
system can dynamically choose which strategy to use for each QPU.
QPUs expose an interface that allows corresponding QPU located in other DCs to
subscribe to their updates.
Each QPU is initially updated by receiving write operations from the storage
system's log.
At the same time, it maintains statistics on how frequently write operations
result to updates to its region of the space.
When the selectivity of write operation for an individual QPU is above a
certain threshold, it can subscribe to updates from the corresponding units in
the other DCs and switch to the second maintenance strategy.
Conversely, when updates from other QPUs are frequent, the QPU can switch again
to receiving updates from the write log.

\subsection{Query Processing Unit Implementation}

Since query processing units operate as services, their internal index
implementation may vary as long as they expose the same interface.
Different instances of the system may implement different index structures
depending on the characteristics of the indexed attributes.

A straightforward implementation is to maintain a simple inverted index for each
indexed
attribute.
Attribute values in each index are sorted, and each value points to a posting
list of primary keys of objects that have this value.
Given a query, the QPU performs lookups in the corresponding inverted indexes
and then calculates the intersection of the retrieved lists of keys.

Since indexes are replicated among DCs, they must be able to converge to a
consistent state even when operations are applied in index replicas in a
different order.
This is accomplished by implementing index structures as a composition of
Conflict-Free Replicated Data Types \cite{syn:rep:sh143} (CRDTs).
CRDTs are replicated data types that guarantee convergence of conflicting
operations without the need for application conflict handling.
The use of CRDTs enables index structures to merge updates originating at
different replicas without the need for central synchronisation or explicit
conflict resolution, despite messages being duplicated and reordered.

However, this mechanism is not sufficient for maintaining a replicating index.
Conflicting concurrent updates to an object's secondary attributes will result
in the introduction of false positives in the index.
In the case of the example described in Section \ref{inter_dc_replication}, 
after propagating updates and merging, the indexes in \texttt{DC1} and
\texttt{DC2} will contain both entries \texttt{Attr:A} and \texttt{Attr:B}.
However, the storage system will choose value \texttt{A} or \texttt{B} based on
a strategy such as last-writer wins.

We address this issue by adding a mechanism that checks query results against
the source data and removes false positives.
Moreover, we use a background task that periodically scans indexes and removes
false positives by checking the source data.

\subsection{Bounding Search Result Staleness}

Query processing units receive updates asynchronously and independently from
each other.
As a result, QPUs may be updated at different rates and diverge from the state
of the storage system and from each other.
Search results may therefore be unboundedly stale relatively to the state of
the data store.

To address this issue, we design a mechanism that enables clients to bound the
staleness of their search results.
Using this mechanism, applications can limit the amount of staleness of each
query, according to their needs.
Since acquiring less stale search results requires additional computations,
applications can make a trade-off between query response time (and resource
utilisation in general) and query result freshness.

We model the storage system's log as a list of write operations.
Each operation that is appended to the log is assigned with a unique
monotonically increasing \texttt{LogID}.
QPUs that are responsible for recent updates use vector clocks to maintain
information on their divergence from the state of the storage system and from
each other.
A vector clock \texttt{VC} consists of an entry \texttt{VC\textsubscript{k}}
for each QPU, indicating that \texttt{QPU\textsubscript{k}} has applied all
write operations up to \texttt{VC\textsubscript{k}}.
QPUs periodically exchange vector clocks, by propagating them to QPUs that
maintain connections to them, following the inverse path of QPU network
connections.
When a QPU processes a new write operation, it increments the corresponding
vector clock entry.
When a QPU receives a vector clock, it merges it with its own by calculating
the maximum value for each entry.
Using its vector clock, a QPU can determine a \texttt{LogID} so that every QPU
in its sub network hierarchy has applied all updates up to that \texttt{LogID}.
We call this \texttt{LogID} a stable index snapshot for this set of QPUs.
Additionally, cached index entries are stored along with their vector clocks so
that QPUs can determine their staleness.

When issuing a query, clients provide an additional argument indicating the
desired level of search result staleness.
This argument has the form of discrete staleness levels, ranging from strongly
consistent to unboundedly stale results.

Given a query and the staleness level attribute, QPUs use vector clock
information to compute the stable index snapshot for the entire QPU network;
The lowest \texttt{LogID} up to which every QPU has applied all updates.
Additionally, the system obtains the \texttt{LogID} of the most recently
appended write operation from the log.
The difference between these two values represents the maximum amount of
staleness for any QPU in the network.
Based on these information and the given staleness argument, the system
determines a target \texttt{LogID\textsubscript{t}}.
Any QPU or cache that contributes to the processing of the given query must
contain the effects of all write operation at least up to
\texttt{LogID\textsubscript{t}}.
Based on the given \texttt{LogID\textsubscript{t}} QPUs ignore older cached
entries and pull write operations from the log, or from QPUs located in other
DCs in order to process the required write operations up to the specified limit.
As a result, query result staleness is bounded by
\texttt{LogID\textsubscript{t}}.

\subsection{Future Directions}

\subsubsection{Adaptive Index Construction}

Our system design enables the implementation of an additional mechanism that
will dynamically construct the distributed QPU network in order to optimise its
structure according to query load and attribute value distributions.
Initially, a single query processing unit will be responsible for the entire
attribute value space.
As new objects are stored in the system, when the number of indexed objects in
a QPU reaches a threshold, the unit will expand the network by spawning a
number of new QPUs and assigning each one of them with a sub-part of its region
of the space.
On the other hand, when, due to deletions, the number of objects which a QPU
indexes reaches below a threshold, it can be merged with a neighbouring unit, 
therefore contracting the QPU network.
This allows QPUs to manage index sizes and prevent fragmentation.

This mechanism can also allow the QPU network to adaptively adjust to query
load.
Additional QPUs can be spawned when the query load in a particular region of
the space is high, in order to spread the load more evenly between QPUs.
Conversely, under-utilised QPUs can be merged to reduce maintenance costs.

Moreover, this mechanism can address the need for multi-resolution indexing.
When a part of the value space is queried with higher resolution, a high load
is introduced to the QPUs which are responsible for these region of the space.
As a result, the QPU network will dynamically expand by spawning more QPUs
responsible for these regions.
Spawned QPUs will be responsible for a smaller part of the hyperspace,
resulting to higher resolution indexing in these regions of the space.

\subsubsection{Dynamic Resource Allocation}

Query processing units operate as services and are not bound to physical
servers in the system.
This enables the use of various mechanisms to dynamically adapt the amount of
computation resources available to the system.

Multiple QPUs with low indexing and query processing load can be collocated in
the same physical servers.
At the same time, highly loaded units can migrate to new servers in order to
have more computation resources available to them and balance load between
servers.
Additionally, the system can spawn multiple instances of a QPU, and place each
one of them placed in a different server.

Furthermore, the strategy of attribute value space partitioning enables two
additional mechanisms for adjusting computation resources available to the
system:
\begin{itemize}
  \item QPUs can cover overlapping regions of the space.
  This allows the system to perform load balancing by having a choice of
  multiple QPUs for parts of a given query.
  \item QPUs can dynamically adjust their boundaries and exchange the regions
  of space they are responsible for.
  Dynamic boundary movement allows the system to re-assign a part of the
  space that is assigned to a highly loaded QPU with limited resources, to
  another QPU with more available resources.
\end{itemize}

\subsubsection{Network structure caching}
 
The hierarchical structure of the distributed QPU network ensures that only
local structure knowledge is required in order to route queries through the
network, and no QPU needs to have a view of the entire network structure.
Each QPU needs to maintain the boundaries and mapping to physical servers, for
the QPUs it is connected to.
In order to process a given query, QPUs recursively determine which of the QPUs
they are connected to should contribute to the query processing and forward the
corresponding sub-queries to them.
However, as the network expands and adds more levels, this process can lead to
increased network traffic due to messages between QPUs.

To address this problem, each QPU can maintains a cache of the structure of the
part of the network that is reachable through it.
The QPU's structure cache stores the boundaries of QPUs as well as their mapping
to servers, for a number connection levels.
QPUs periodically send their cache to higher to the network hierarchy, so that
network structure changes are propagated through the network.
Modifications to the network structure, such as additions of new QPUs and
boundary adjustment are performed locally and then propagated upwards.
Using this mechanism, QPUs can decompose and route queries without the need to
go through the entire structure of the network, reducing thus the number of
messages needed to sent thought the network to process each query.

However, since caches are updated asynchronously, there may be cache misses when
sub-queries are forwarded to units that no longer exist or are not placed in
the destination servers.
When a cache miss occurs, QPUs iteratively backtrack and use higher levels of
the network hierarchy until the query can be successfully processed.

\section{Discussion and Next Steps}

We have introduced and analysed the research problem of extending
geo-distributed, weakly consistent data stores to provide real-time search on
secondary attributes.
We have presented our study of the state-of-the-art on various fields related
to providing secondary attribute search in distributed systems.

Our literature review shows that secondary indexing systems are commonly 
designed based on specific system architecture and workload characteristics.
This shows that no optimal design exists for an indexing and query processing
system, and implementing such systems is based on making trade-offs according to
the target use case characteristics.
Based on this observation, we have analysed the design space and discussed how
various design choices affect the efficiency of the system and address the
problem's requirements.

Moreover, most related works focus more on the low level design of indexing data
structures and their maintenance, and less on the distributed nature of the
system.
To our knowledge, there is no work considering an indexing system that is
replicated among multiple data centres.
Moreover, most approaches choose to maintain indexes that are strongly
consistent with the source data by updating index structures synchronously, in
the critical path of write operations.
We have introduced and described our approach which focuses on the
geo-distributed nature of the system, and explores the space between strong
and eventual index consistency.

As a first step, we have proposed the mechanisms that can be used for
implementing a system that efficiently addresses our problem statement.
At next steps of our work, we intend to define the algorithms and policies that
will make use of the described mechanisms to implement the system's
functionality, and implement a prototype of the system.

We plan to use greedy algorithms and heuristics for processing queries using
the distributed network of query processing units instead of using query
optimization techniques.
We believe that applying query optimization techniques in order to find the
optimal processing strategy for each query is out of the scope of our approach,
and we should instead use best effort techniques which minimize response time.
As future work, we intend to apply learning algorithms to tune the parameters of
our greedy algorithms and optimise them at runtime.

Furthermore, our model of QPUs that work as services and are not bound to
specific system servers enables us to expand the QPU network towards the edge of
the network, and place QPUs to client machines.
Performing computations at the edge enables the system to reduce the computation
load at the code of the system, enable more sophisticated indexing techniques,
and improve availability in case of network partitioning.

\bibliographystyle{unsrt}
\bibliography{report}

\end{document}